\def\year{2022}\relax
\documentclass[letterpaper]{article} 
\usepackage{aaai22}  
\usepackage{times}  
\usepackage{helvet}  
\usepackage{courier}  
\usepackage[hyphens]{url}  
\usepackage{graphicx} 
\usepackage{natbib}  
\usepackage{caption} 
\DeclareCaptionStyle{ruled}{labelfont=normalfont,labelsep=colon,strut=off} 
\pdfoutput=1
\usepackage{balance}
\usepackage{wasysym}
\usepackage{xcolor}         

\usepackage{enumitem}
\usepackage{mathptmx}
\usepackage{graphicx}
\usepackage{array}
\usepackage{multirow}
\usepackage{pifont}
\usepackage{color, colortbl}
\usepackage{siunitx}
\usepackage{booktabs}
\usepackage{rotating}
\usepackage{xparse}
\usepackage{tabularx}
\usepackage{url}
\usepackage{amssymb}
\usepackage{pifont}
\usepackage[font=small]{caption}
\usepackage{subcaption}
\usepackage{tikz}
\usepackage{makecell}
\usepackage[ruled,vlined]{algorithm2e}
\usepackage{float}
\usepackage{verbatim}
\usepackage{soul}
\usepackage{xspace}

\pdfoutput=1
\renewcommand\paragraph[1]{\vspace{3pt} \noindent\textbf{#1.}\hspace{0.5em}}

\begin{document}
\title{No Calm in The Storm: Investigating QAnon Website Relationships\thanks{Paper accepted at ICWSM 2022. Please cite the ICWSM version.}}
\author {
    Hans W.A. Hanley, 
    Deepak Kumar, 
    Zakir Durumeric\\
}
\affiliations {
    Stanford University \\
    hhanley@stanford.edu, kumarde@stanford.edu, zakird@stanford.edu
}
\maketitle
\definecolor{Gray}{gray}{0.9}
\begin{abstract}
\begin{quote}
QAnon is a far-right conspiracy theory whose followers largely organize online. In this work, we use web crawls seeded from two of the largest QAnon hotbeds on the Internet, Voat and 8kun, to build a QAnon-centered domain-based hyperlink graph. We use this graph to identify, understand, and learn about the set of websites that spread QAnon content online. Specifically, we curate the largest list of QAnon centered websites to date, from which we document the types of QAnon sites, their hosting providers, as well as their popularity. We further analyze QAnon websites' connection to mainstream news and misinformation online, highlighting the outsized role misinformation websites play in spreading the conspiracy. Finally, we leverage the observed relationship between QAnon and misinformation sites to build a highly accurate random forest classifier that distinguishes between misinformation and authentic news sites. Our results demonstrate new and effective ways to study the growing presence of conspiracy theories and misinformation on the Internet.

%
%
\end{quote}

\end{abstract}

\section{Introduction}
QAnon is a far-right conspiracy that started in 2017, growing from a fringe Internet theory into a massive political movement during Donald Trump's presidency. QAnon alleges that a cabal of Satanic, cannibalistic pedophiles control American politics and media. The movement has a proclivity towards violence, which peaked on January 6, 2021, when its followers joined in the insurrection attempt at the U.S.\ Capital~\cite{Ovide2021}. QAnon supporters largely organize online, using both mainstream and alternative platforms to discuss the conspiracy and to organize real-world events~\cite{Kidnap2020}. In response to QAnon's growth, many Internet companies, including Google, Twitter, Facebook, and Reddit have taken steps to remove QAnon content from their platforms~\cite{Greenspan2020}. Yet, despite these bans, QAnon remains a prominent movement.

In this work, we utilize web crawling and the resultant hyperlink graph to shed light on the online QAnon ecosystem. We analyze the extent of the conspiracy theory's spread, its most important platforms and websites, and its connection to both mainstream media and misinformation sites. We also show how the relationships between QAnon websites and other domains can be utilized to accurately identify misinformation more broadly.

We begin our study by curating the largest known corpus of QAnon websites. To do this, we crawl two hotbeds of QAnon content, 8kun and Voat, as well as their hyperlinked URLs to develop a directed hyperlink graph of candidate-QAnon websites. Utilizing the intuition that websites typically hyperlink to semantically similar sets of websites, the Szymkiewicz--Simpson coefficient graph similarity statistic, and manual confirmation, we curate a final set of 324~unique QAnon dedicated websites. We develop a taxonomy of these QAnon websites, catalog their role in spreading the conspiracy theory, and document their popularity and hosting providers.

Next, we investigate the relationships between QAnon websites and other Internet platforms using Hypertext Induced Topic Selection ({HITS}) analysis. We show that despite bans, Twitter and Reddit continue to link to QAnon content and remain prominent \emph{network hubs} in the QAnon ecosystem. Using this same analysis, we find strong ties between QAnon websites, misinformation sites, and platforms known for hosting abusive content. These websites include Free Republic, American Thinker, and BitChute, a popular video hosting platform described by the Southern Poverty Law Center as filled with ``hate-fueled material''~\cite{Hayden2019}.

Finally, we show that there is a strong connection between QAnon websites and authentic and misinformation news sources. While QAnon sites regularly link to both types of content, typically only misinformation websites (e.g., {americanthinker.com}, {freepublic.com}, and {thegatewaypundit.com}) directly reference QAnon sites, adding to the wealth of evidence that sites like these contribute to toxic political ``echo-chambers''~\cite{starbird2018ecosystem}. Building on this observation, we demonstrate how features derived from our QAnon-based hyperlink graph can be used to identify misinformation more broadly. We train a random forest classifier on hyperlink connections to QAnon sites to label domains as misinformation or authentic news. With this model, we achieve 0.98~AUC on a 30\% test set, illustrating how the relationships between websites and QAnon domains can be used to understand misinformation and conspiracy theories generally.
While deplatforming QAnon from social media platforms may be beneficial in reducing its spread, the decision simultaneously increases the barrier to understanding misinformation. Our work shows how web crawling and hyperlink graph analysis can lead to better a understanding of conspiracies and online misinformation. We hope that our results help platforms curb the spread of QAnon as well as provide a robust methodology for studying misinformation moving forward.


%


\section{Background}
QAnon is a conspiracy theory premised on the notion that the world is run by a cabal of pedophiles who Donald Trump will purge in a day of reckoning known as ``The Storm''~\cite{papasavva2020qoincidence}. The movement began in October~2017 when an anonymous 4chan user ``Q'' asserted that he was an official in the United States Department of Energy with high (Q-level) clearance, outlining the conspiracy in a post titled ``Calm Before The Storm.'' Q asserted that he was sending the world through a ``Great Awakening'', and since then has regularly posted on 8kun, where Q's followers try to uncover the meaning in his messages known as ``Q drops.'' Because these followers post \textbf{anon}ymously, the conspiracy theory is known as QAnon (``Q + Anon'').

QAnon has grown into a massive political movement in the United States, with an estimated 56\% of self-identified Republicans believing that QAnon theories are at least partly true according to a Civiqs poll.\footnote{https://civiqs.com/results/qanon\_support} More troubling than its popularity is the theory's tendency to incite violence. On January 6th, 2021, the movement's power was shown in full force when QAnon followers joined other conspiracy theorists to storm the U.S. Capitol~\cite{Ovide2021}. As a result of its violent tendencies, the United States Federal Bureau of Investigation (FBI) considers QAnon a domestic terror threat and QAnon followers as domestic extremists~\cite{Winter2019}.

QAnon's growth is largely attributed to online social media. Between April and August 2020, Facebook posts about QAnon nearly doubled from 992 to 1,772~posts per day~\cite{FacebookQAnon}. By August 2020, over 3M~Facebook accounts subscribed to QAnon pages, and 157K~Twitter accounts were actively spreading QAnon information~\cite{fbmembers,Timber2020}. Since then, several social media companies and technology platforms have attempted to curb QAnon's influence by ``deplatforming'' the group. In 2018, Reddit banned subreddits devoted to QAnon, proclaiming Reddit QAnon-free in September 2020~\cite{Tiffany2020}. In July 2020, Twitter stopped recommending accounts and content on 150K accounts that shared QAnon material, suspending 7K outright~\cite{twitter-suspends-more}. By October 2020, Facebook terminated over 1,500 pages, groups, and profiles dedicated to QAnon~\cite{FacebookQAnon}. Other platforms including Google, Triller, YouTube, Etsy, Pinterest, Twitch, Discord, Spotify, Vimeo, Patreon, and even the fitness company Peloton have taken steps to eliminate the movement from their platforms~\cite{Greenspan2020}. However, despite the bans, QAnon remains a prominent and dangerous movement. This is in part because QAnon thrived throughout 2020 in two digital homes:

\paragraph{8kun}
{8kun.top} (previously 8chan) is an imageboard that prides itself on minimal moderation~\cite{Glaser2019}. New ``Q drops'' are posted on 8kun, with the site acting as the locus of the QAnon conspiracy theory. The main board dedicated to QAnon on 8kun is Qresearch; several prominent offshoots also exist including Qpatriotresearch, Qresearch2gen, Abcu, Qsocial, and Qpatriotswwg. Prior to hosting QAnon content, 8kun was used to spread the manifestos of the Christchurch, Poway synagogue, and El Paso shooters~\cite{Glaser2019}. This resulted in Google removing 8kun from search results and Cloudflare, a major CDN, terminating service for the site. 8kun became publicly accessible again in late 2019 after the Vancouver, WA based Internet provider Vanwatech agreed to host it~\cite{Krebs2021}.\looseness=-1

\paragraph{Voat}
{Voat.co} (previously WhoaVerse) was a news aggregator/Reddit alternative founded in 2014 by Atif Colo that shut down on December 25, 2020~\cite{Robertson2020}. Voat had a self-described focus on free speech but was known to host ``extremist right-wing content''~\cite{Robertson2020}. Content on Voat was organized into areas of interest called ``subverses'', which were similar to subreddits. Voat received an influx of users following Reddit banning the PizzaGate conspiracy theory in 2016 and again in 2018 when Reddit banned all QAnon content~\cite{papasavva2020qoincidence}. As a result, and as referenced on 8kun, {voat.co/QRV} became the ``official'' home of QAnon research. Many other subverses were also dedicated to Q content: v/QAnon, v/GreatAwakening, v/TheGreatAwakening, v/CalmBeforeTheStorm, v/PatriotsSoapbox, v/QRV, and v/theawakening~\cite{papasavva2020qoincidence}. Like 8kun's imageboard, Voat provided an easy means for followers to interact and discuss QAnon theories.\looseness=-1

\section{Curating a List of QAnon Websites} 
We begin our study by identifying and analyzing websites dedicated to QAnon. We detail how we leverage links posted on 8kun and Voat to build a resulting hyperlink graph containing 324~unique QAnon websites. We then analyze the types, popularity, and domain hosting of the websites dedicated to the conspiracy theory.

\subsection{Building from QAnon Seed Sites}
We use 8kun and Voat as our initial QAnon seed sites. To extend this set of QAnon sites, we first collect the sites linked by QAnon forums/subverses on 8kun and Voat. On 8kun, we identified 6~QAnon subcommunities by inspecting the 413~board descriptions on the site, manually checking those that claimed to discuss QAnon. For Voat, we compiled a list of 18~QAnon subverses based on the list curated by Papasavva et~al.~\cite{papasavva2020qoincidence}. Voat submissions are ephemeral, which limited our analysis to URLs available between August and December 2020 (on December 25, 2020, Voat permanently shut down). For both 8kun and Voat, a handful of popular pages contributed most of the hyperlinks (e.g., Qresearch on 8kun and QRV on Voat), but we include the full set of QAnon forums/pages for maximal coverage of QAnon-related material. In total, we collected 4M~unique URLs: 3.9M from 8kun and 160K from Voat. These URLs belong to 6.9K and 7.2K~domains, respectively. We refer to this set of URLs from Voat and 8kun as 8kun's and Voat's \emph{Hop 1 Links}.

For each of 8kun's and Voat's \emph{Hop 1 Links}, we fetch the URL, parse the DOM tree, and collect the hyperlinks to other pages (i.e., HTML \texttt{<a>} tags). We then crawl the external sites linked to by the \emph{Hop 1 Links}, creating an extended set of \emph{Hop 2 Links}. From these two hops, we generate a directed hyperlink graph of domains that link to one another. 

\paragraph{Ethical Considerations}
We collect only publicly available data on Voat and 8kun. We use the Voat API to crawl data from their website; our data collection abided by Voat's terms of service. We crawled 8kun's website; 8kun has no terms of service. We did not attempt to deanonymize any Internet user in the study. We further follow the best practices for web crawling as in works like Acar et al.~\cite{acar2014web}.

\subsection{Finding New QAnon Domains} 

While analyzing the sites within 8kun's and Voat's \textit{Hop 1 Links}, we found that there were relatively few domains that hyperlinked back to 8kun and Voat: 80 out of 6.9K sites for 8kun and 112 out of 7.2K for Voat (180 unique domains). Upon manual inspection, we found that a significant fraction of these bidirectionally-connected websites were Qanon-focused (47 out of 80~bidirectional links for 8kun; and 51 out of 112~bidirectional links for Voat). We classify a website as a \textbf{QAnon domain} if the site has a QAnon banner on its homepage, has forums for promoting and discussing QAnon theories, contains a repository of Q ``drops'', sells QAnon merchandise, or is a social media site made for QAnon followers. 

We also found that our set of manually-verified QAnon sites often hyperlinked to other QAnon websites beyond Voat and 8kun, hinting that we could find additional QAnon websites by analyzing a larger hyperlink graph structure. We extend our initial hyperlink graph by more deeply crawling the 180~domains with bidirectional links with Voat and 8kun. We then used a node similarly metric to predict if a domain is Qanon-focused before performing manual verification. We detail our full crawling and graph analysis methodology below:

\paragraph{Crawling Methodology} Realizing that we could not scrape every page on every domain that we collected, we developed a methodology to scrape a large subset of each domain's pages. To crawl a website, we iteratively retrieve the internal pages that are up to 15~hops away from the domain's root page. For each site, we thus fetch the root page, parse the DOM tree, and collect any additional links to internal pages, and then iteratively repeat the process to create a new set of \emph{Hop 1 Links} for each website. We show our algorithm, DEEPCRAWL, in Algorithm~\ref{alg:crawl}. We then collect the hyperlinks to external sites from each of the pages of our new \emph{Hop 1 Links}, creating an extended set of new \emph{Hop 2 Links}, from which we generate a directed graph of domains that connect to one another. 

\begin{algorithm}
\SetAlgoLined
\KwResult{List of URLs from given domain}
 pagesToCrawl = domain homepage\;
 newPagesToCrawl = empty set\;
 crawlSet = empty set\;
 collectedURLs = empty set\;
 i = 0\;
 \While{ i $< 15$ and size(pagesToCrawl) $> 0$ }{
  \For{page in pagesToCrawl} {
     urls = scrape(page)\;
     crawlSet.append(url)\;
    \For{url in urls} {
        \If {url not in crawlSet and url.domain$==$domain } {
            newPagesToCrawl.append(url)\;
        }
        \Else{ 
        	 crawlSet.append(url)\;
         }
    }
  }
   $i\gets i+1$\;
   pagesToCrawl $\gets$ newPagesToCrawl\;
  newPagesToCrawl $\gets$ nil\;  
 }
 \Return crawlSet\;
 \caption{DEEPCRAWL(domain)}
 \label{alg:crawl}
 \end{algorithm}
 
\begin{figure*}
\centering
\includegraphics[width=2\columnwidth]{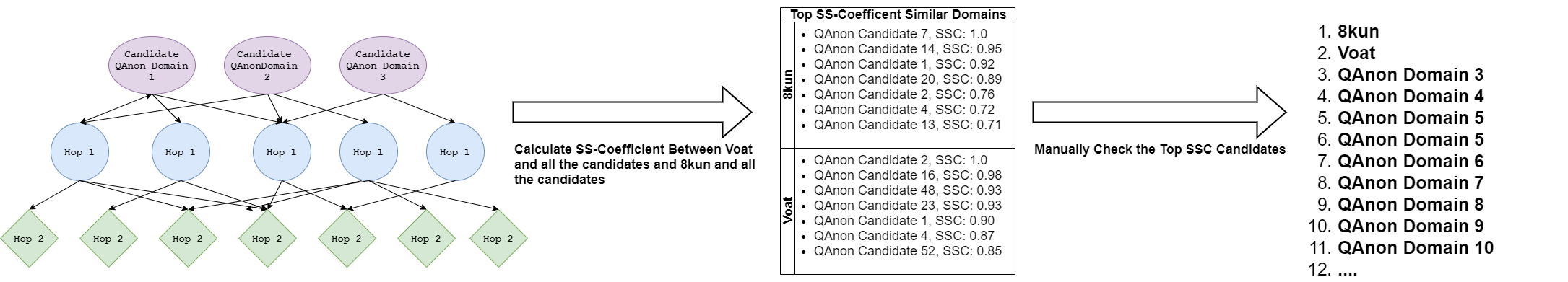}
\caption{\textbf{Illustration Of Finding New QAnon Domains}---We utilize DEEPCRAWL and the Szymkiewicz-Simpson coefficient between candidate domains and known QAnon domains to build a larger list of QAnon-focused domains.
}
\label{fig:finding-q}
\vspace{-10pt}
\end{figure*}

\paragraph{Computing Similarity Based on Links}
In our initial set of 180~sites with links to 8kun and Voat, we observed that 88~were centered around the QAnon conspiracy while 92~were not. Now, with the intuition that similar sites often hyperlink to similar sites (\textit{e.g.} QAnon to QAnon), in order to help identify candidate QAnon websites in our new crawls, we wanted to understand if we could use graph metrics to help identify QAnon-focused website candidates. We thus utilize the Szymkiewicz-Simpson coefficient (SSC) similarity metric to assess how similar candidate websites are to Voat and 8kun, and thus Qanon-focused. SSC is a metric used to measure the similarity of nodes in a graph using the overlap of their connections. The SSC for two sites $x$ and $y$ is computed as follows:

\begin{equation}
    SSC(X,Y) = \frac{| X \cap Y | }{\min(|X|,|Y|)}
\end{equation}

\noindent
where X is the set of sites with connections with domain $x$, and Y is the set of sites with connections to domain $y$. We thus calculate the SSC of each of the 180~bidirectionally linked domains relative to 8kun and Voat to understand if there is a difference in SSC value between QAnon-focused and non-Qanon-focused websites.

QAnon domains have a statistically significant difference in SSC values with 8kun compared to non-QAnon domains (Mann-Whitney U statistic 1583.5, $p=2.2\times10^{-8}$). The QAnon domains in our first crawl have an average SSC of 0.5 with 8kun compared to an average SSC of 0.33 for non-QAnon websites. Similarly, we find the QAnon domains have an average SSC with Voat of 0.52 while the non-QAnon sites have an average SCC of 0.37 (Mann-Whitney U statistic 1750, $p=4.8\times10^{-7}$). We thus find we can utilize SSC to filter a list of candidate websites to a size that can be subsequently manually verified. We detail our graph creation and QAnon identification techniques in Figure~\ref{fig:finding-q}.

\paragraph{Final QAnon Website List}
Using our list of 90~confirmed QAnon domains (88 new QAnon domains + Voat and 8kun) and the corresponding hyperlink graph, we identified an additional 3K~websites that had bidirectional links with this set of 90~known QAnon domains. After performing DEEPCRAWL on each of the 3K~websites, we narrowed our list of QAnon candidate websites to the top 10~most SSC-similar websites to each of the QAnon domains in our list of 90 (unique total of 600 websites). SSC thus enabled us to narrow down the list of 3K~websites that had bidirectional links with our set of 90~QAnon domains to 600~candidate sites. We manually checked these 600~sites for QAnon content, filtering down to a final set of 324~QAnon websites.

\subsection{Types of QAnon Websites}
QAnon websites come in five broad categories. For each QAnon website that appears in the Alexa Top Million list of popular websites, when mentioned, we give its rank as of January 1, 2021~\cite{amazon-top-mil}.  



\paragraph{Q Drop Sites} We identify 29~sites that store official ``Q drops'' from 8kun in an easy-to-read format. These sites typically allow users to view each ``drop'' in chronological order. As noted previously, Q drops are cryptic messages that contain vague references to current events or famous people. For example, Q drops often reference former President Trump's Twitter account or famous individuals like Bill Clinton and General Michael Flynn. Our list includes sites like {qanon.pub} (Alexa rank 27945), {qdrop.pub}, and {operationq.pub}.

\paragraph{Q News and Research Sites} We find 196~Q news and research sites where users can download and interact with QAnon content. Research sites normally have open discussions of Q (one even provides a spreadsheet that allows users to post questions about the QAnon conspiracy). Other research sites merely present analyses of the news and how events fit within the QAnon conspiracy theory. Our list includes sites like {x22report.com} (Alexa rank 12,573), {wwg1wga.martingeddes.com}, and {theqpatriothub.weebly.com}.

\paragraph{QAnon Merchandise and Stores} We uncover 23~sites that produce QAnon entertainment, sell QAnon merchandise, or otherwise profit from the conspiracy. These websites produce merchandise ranging from books, podcasts, and movies to commemorative coins, trading cards, and branded coffee. Additionally, two sites actively promote real-life QAnon conferences where QAnon followers can meet in person. Our list includes {qcon.live}, {thegreatawakeningsummit.com}, and {thebookofqanon.com}.

\paragraph{QAnon Social Media Clones} In addition to websites that promote the QAnon conspiracy theory with blog posts and ``news'' articles, we found 14~QAnon domains that operate as social media sites. Several of these sites are simply QAnon-branded versions of popular sites like Reddit, Facebook, or Twitter. Our list includes sites like {voat.co} (Alexa rank 64,488), {wg1wga.com}, and {anonup.com}.

\paragraph{Non-U.S. QAnon Sites} Despite QAnon being a U.S.-focused conspiracy, we note our list includes 62~non-U.S. focused QAnon sites that export the conspiracy theory abroad, consistent with recent research and reporting~\cite{Karlssson2020,Farivar2020}. We find websites targeted toward audiences in Germany (22), France (13), Japan (5), the Netherlands (4), Italy (3), the UK (3), Belgium (2), New Zealand (2), Sweden (2), Canada (1), Australia (1), and Switzerland (1). We also find 3~domains directed at Spanish-speaking audiences. We lastly note that ``Q drops'' were translated into English, Dutch, French, German, Italian, Korean, Portuguese, Spanish, and Swedish on one of the {Q Drop sites}. Some examples of non-U.S. QAnon websites include {pit-hinterdenkulissen.blogspot.com} (Alexa rank 38,549), {digital-soldier.de}, and {wwg1wgadutch.wordpress.com}.

\begin{figure}
\centering
\includegraphics[width=\columnwidth]{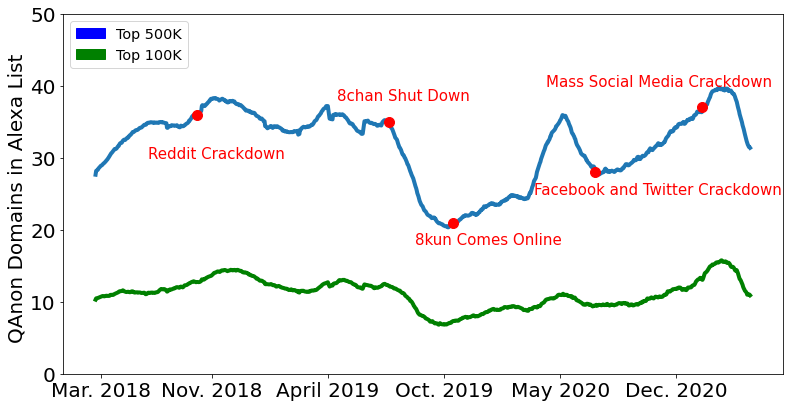}
\caption{\textbf{QAnon Domains In Amazon Alexa Top Lists}---QAnon domains have consistently appeared in the Alexa Top Million Websites List. We present the dates when different social media companies announced they had taken action against the QAnon movement.}
\label{figure:qanon-popularity}
\vspace{-10pt}
\end{figure}

\subsection{QAnon Site Popularity}

Several QAnon sites are relatively popular according to the Alexa Top Million list~\cite{amazon-top-mil}. While we do not perform any causal analysis, we observe several coarse trends in the popularity of QAnon over time. From the movement's inception until now, there have been several crests (the Spring of 2020) and troughs (following the shutdown of 8chan in the Autumn of 2019), with interest in the movement peaking in the aftermath of the January 6th U.S. Capitol attack. Figure~\ref{figure:qanon-popularity} shows the number of domains in the Alexa Top 100K and 500K lists along with the dates when different social media companies \emph{announced} they had taken action against QAnon accounts (this explains why the Facebook and Twitter crackdowns occur after a peak in QAnon interest). The largest decline in popularity was in August 2019, when Cloudflare dropped 8chan (now 8kun) after a mass shooting in El Paso where the shooter explicitly referenced inspiration from 8chan~\cite{Price2019}. QAnon sites did not regain popularity until after 8chan was rebranded as 8kun, migrated to Vanwatech, and became protected by DDoS-Guard, ``a dodgy Russian firm that also hosts the official site for the terrorist group Hamas''~\cite{Krebs2021}. 
We lastly note that QAnon domain registrations have increased steadily since 2017; many popular QAnon domains are newly registered. While only 6~sites were registered in 2017, 34~were in 2018, 40~were in 2019, and 51~were in 2020. The rest of the websites in our list are older sites that eventually shifted their focus to QAnon.

\subsection{QAnon Website Hosting} 

The 324~QAnon sites we identify are hosted by 75~providers, most prominently Automattic (58 domains), Google (53 domains), Cloudflare (39 domains), GoDaddy (18 domains, 11\%), and Vanwatech (11 domains). Automattic, the home of WordPress, appears here due to the many WordPress blogs promoting QAnon content. 11~QAnon domains are hosted by the relatively unknown provider Vanwatech, which rose to prominence by offering security services and a decentralized content delivery network to 8kun after Cloudflare dropped it~\cite{Krebs2021}. Despite dropping 8kun in 2019, Cloudflare and Google still host 17.9\% and 12.0\% of QAnon domains respectively, highlighting their role in propping up the conspiracy online.

\section{The Full QAnon Ecosystem}
\begin{figure}
\centering
\includegraphics[width=\columnwidth]{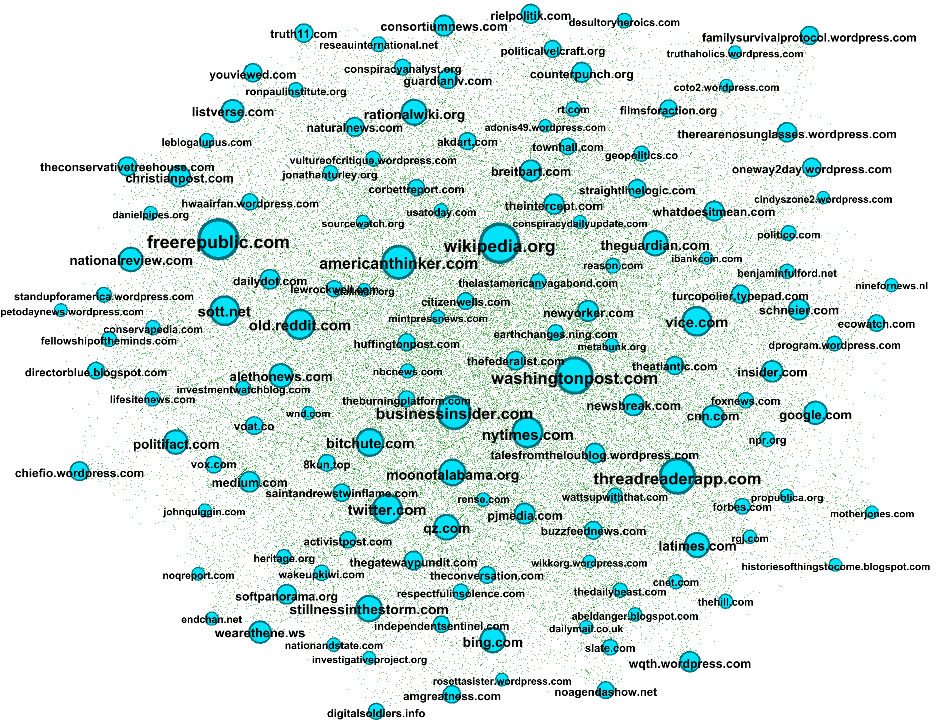}
\caption{\textbf{QAnon Cluster Hubs}---Top hubs surrounding QAnon domains as determined by the hub centrality metric.}
\label{fig:tophubs}
\end{figure}

In this section, we analyze the graph of known QAnon websites to identify prominent hubs in the QAnon ecosystem. We further illuminate QAnon's connection to online communities, including social media platforms, authentic mainstream news providers, and known misinformation sites.

\subsection{QAnon Ecosystem Hubs}

To understand the structure of the QAnon ecosystem, we build a hyperlink graph centered around QAnon using the {Hypertext Induced Topic Selection} ({HITS}) algorithm~\cite{Kleinberg1997}. Starting with the 324~QAnon domains documented earlier as our \textit{root set} of domains, we build a \textit{base set} of domains by collecting the domains that both hyperlink to or are hyperlinked by these QAnon websites. In the directed graph of domains connected to QAnon websites, there are 198K~domains and 3.6M~directed edge connections. 
\begin{table}
\centering
\small
\begin{tabular}{lrrr}
\toprule
{Domain} & {Hub} & {Authority}  & {Alexa Rank}  \\
\midrule
freerepublic.com               &   0.135    &   0.022   &   27,045 \\
wikipedia.org                &   0.132    &   0.054   &   13  \\
washingtonpost.com     &   0.122    &    0.054  &   163  \\
threadreaderapp.com                    &  0.119    &   0.017   &   23,836  \\
businessinsider.com                    &  0.111    &   0.044   &   227  \\
americanthinker.com             &   0.111   &   0.028   &   12,964 \\
nytimes.com                &  0.101   &   0.057   &   88 \\
reddit.com                   &   0.099    &   0.007   &   17 \\
twitter.com    &    0.098     & 0.081   &   42  \\
vice.com    &   0.097    &   0.032   &   24,909  \\
sott.net             &   0.097    &  0.020    &   71,876 \\
bitchute.com         &   0.090    &  0.021    &   1,522 \\
\bottomrule
\end{tabular}
\caption{\textbf{Top Hubs in the QAnon Cluster}---The top 10 domain hubs surrounding our QAnon cluster with their normalized [0,1] hub and authority centralities. Alexa rank from January 1, 2021.}
\vspace{-10pt}
\label{table:qanon_hubs}
\end{table}
Per {HITS}, we discern both \textit{hub} and \textit{authority} central domains~\cite{LinkAnalysis}. Hub domains are defined as domains that point to many ``high authority domains and which are themselves high-value domains''. A hub typically has a large out-degree. Authority domains are defined as ``valuable and information-dense domains'' pointed to by many other domains. An authority ordinarily has a high in-degree. Due to the nature of the authority metric, authorities tend to be globally popular domains. In our analysis the highest authority domains are {youtube.com}, {twitter.com}, {facebook.com}, and {google.com}. Given this pattern, we primarily focus on the hubs in the QAnon ecosystem (Table~\ref{table:qanon_hubs}). 


\paragraph{Social Media and Platform Hubs}
Several prominent mainstream social media sites are hubs, most notably Twitter and Reddit. 
Twitter is the ninth-largest hub overall; Twitter links to 113~QAnon domains and is linked to by 248~QAnon domains. Reddit similarly links to 23~QAnon domains and is hyperlinked to by 128~different QAnon domains in our crawl. Another significant hub, YouTube, has links to 23~QAnon domains and 239~QAnon websites link to it.  Last, we note {threadreaderapp.com}'s central role within the QAnon ecosystem. Thread Reader App is a service that allows users to view Twitter threads more seamlessly. A subset of Twitter, 80~QAnon domains link to Thread Reader App pages while it links to 92~different QAnon websites. These results indicate that while social media platforms are trying to remove QAnon content, they still serve as hubs of QAnon content and enable Internet users to access the conspiracy theory.

\paragraph{Authentic News Media}
We define \emph{authentic news} as news websites that are labeled as ``reliable'' or simply ``biased'' by OpenSources\footnote{https://github.com/several27/FakeNewsCorpus}, PolitiFact\footnote{https://www.politifact.com/article/2017/apr/20/politifacts-guide-fake-news-websites-and-what-they/}, Snopes\footnote{https://github.com/Aloisius/fake-news}, and Mellissa Zimdars\footnote{https://library.athenstech.edu/fake}. These websites generally adhere to journalistic norms, have attributed authors, and avoid publishing outright false claims. We expand the category by additionally including the websites of the 61 Pulitzer Prize-winning newspapers going back to 1981.

Several mainstream news sources, including the New York Times, Washington Post, and Business Insider are prominent hubs in the QAnon network (Table~\ref{table:qanon_hubs}). All three occasionally link to QAnon websites like {8kun.top} in news articles. QAnon domains furthermore frequently link to these news sites: 139~QAnon domains link to The New York Times, 133~link to the Washington Post, and 112~link to Business Insider. We find that QAnon sites frequently link to mainstream sources, ridiculing them as ``fake news'' or using their news stories (e.g., presidential election news) to prop up their theories (see Figure~\ref{figure:qpost}). 

\paragraph{Misinformation and Alternative News Media}We define \emph{misinformation} as websites labeled as ``fake-news'' or ``disinformation'' in lists curated by OpenSources, PolitiFact, Snopes, and Melissa Zimdars. 

As seen in Figure~\ref{fig:tophubs} and Table~\ref{table:qanon_hubs}, several misinformation websites including the Free Republic, the American Thinker, and sott.net play prominent roles. A total of 44~QAnon domains have links to the Free Republic, 83 to the American Thinker, and 70 to Sott.net. We observe that while QAnon websites utilize authentic news media as a foil, they simultaneously promote several misinformation-spreading websites to reinforce their argument (see Figure~\ref{figure:qpost_Con}). 

\paragraph{QAnon websites}
While not in the top 10~largest hubs in the ecosystem, several QAnon domains play outsized roles in our analysis. The largest include {stillnessinthestorm.com} (hub centrality 0.086), {wqth.wordpress.com} (0.065), and {voat.co} (0.057), and {8kun.top} (0.053). {stillnessinthestorm.com} is a blog/news site that posts regularly about QAnon and other conspiracy theories. Others including the Columbia Journalism Review\footnote{https://www.cjr.org/fake-beta} have already documented its role in spreading misinformation. {wqth.wordpress.com} is a personal blog that seeks to interpret how current events fit into the QAnon conspiracy theory. Due to concern over potential deplatforming, the site recently moved off of WordPress to the independent domain {theqtree.com}.\looseness=-1

\paragraph{BitChute} The video hosting platform BitChute is also one of the
largest hubs in our graph. 152~QAnon domains link to BitChute videos, highlighting its role in spreading the conspiracy. In our initial scrape, we also found that BitChute links to at least 132~QAnon domains. Twitter began blocking links to {bitchute.com} in August 2020 (Hinchcliffe 2020) due to extreme content on the site~\cite{Hinchcliffe2020}.

\paragraph{Wikipedia} 163~QAnon domains link to Wikipedia, while Wikipedia links to 11~QAnon websites in our crawl. Many QAnon domains use Wikipedia in order to provide definitions and give background to their arguments. For example, in referring to the ``mainstream media'' as traitors or a \textit{fifth column}, one QAnon follower on the GreatAwakening Voat.co subverse linked to a Wikipedia article defining the term.

\subsection{QAnon's Connection to News Outlets}
\begin{figure}
\centering
\includegraphics[width=\columnwidth]{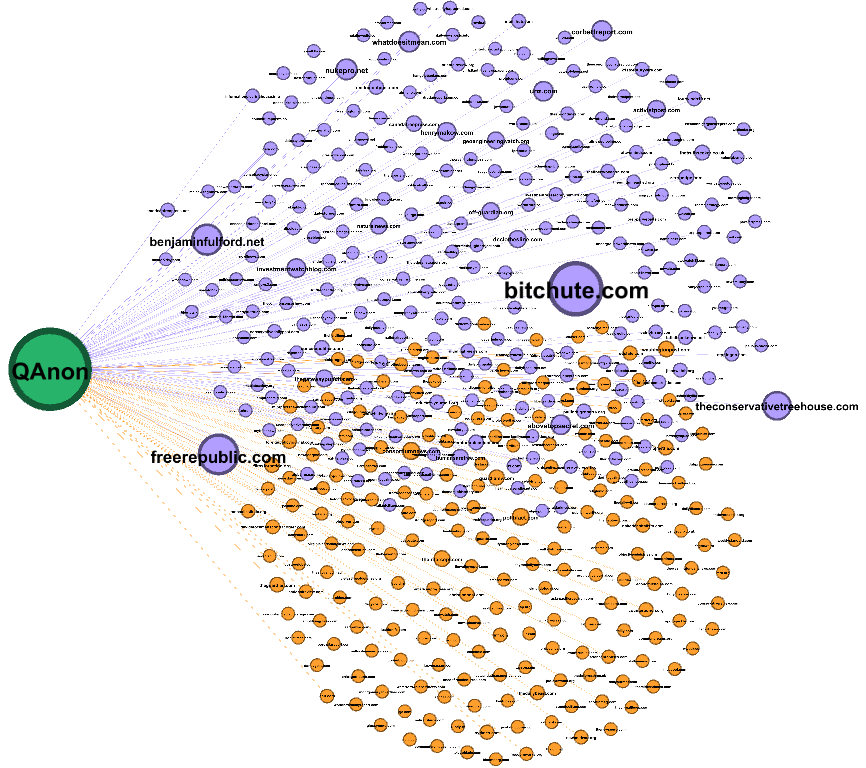}
\caption{\textbf{Websites That Point To  QAnon Domains}---%
Graph of connections directed from authentic news websites (orange) and misinformation websites (purple) to QAnon domains. The size of domain is determined by the number of QAnon domains it points to. As seen while both authentic and misinformation websites have directed connections with QAnon, the websites with the most connections are largely misinformation sites. 
}
\label{figure:website-to-qanon}
\vspace{-10pt}
\end{figure}
In our initial hub analysis, we found that several popular news outlets including the New York Times, Washington Post, Business Insider, and the American Thinker play prominent roles as hubs. Given QAnon's relationship with political events of the last four years, this deep connection is not surprising. We now more broadly quantify the relationship between QAnon and both {misinformation} news sources and {authentic} news sites. To gain a firmer understanding of these relationships, we additionally DEEPCRAWL 167~authentic news sites and 189~known misinformation sites to expand our analysis.

\paragraph{Authentic News Sites} We select 167~authentic news websites from our list of news websites labeled as  “reliable”  or simply  “biased” by OpenSources, PolitiFact, Snopes, and Mellissa Zimdars. We crawl these 167~authentic news websites from January to April 2021, including the sites belonging to the New York Times, Washington Post, and The Guardian. This set of authentic news sites has connections to 320K~other websites in our hyperlink graph.

\paragraph{Misinformation Sites}We select 189~misinformation websites from our list of news websites labeled as  “fake-news”  or “disinformation” by OpenSources, PolitiFact, Snopes, and Mellissa Zimdars. We crawl these 189~misinformation domains from February to April 2021. Examples of misinformation domains include Breitbart, Globalresearch, and Unz, all of which have played a role in spreading conspiracy theories and misinformation online~\cite{starbird2018ecosystem}. This set of misinformation domains has connections with 202K~other domains in our hyperlink graph.


\paragraph{Relationship to Authentic Sources}
Relatively few authentic media sites have connections with QAnon domains. After crawling our list of 167~authentic news sites, we found that only 36 (22\%) hyperlink to QAnon sites and that {authentic} news websites hyperlink to an average of only 0.8~QAnon sites. The most prominent {authentic} news websites with directed connections to QAnon are consortiumnews.com (14~QAnon domains) and politifact.com (13~QAnon domains). This paucity of connections can be seen most directly in Figure~\ref{figure:website-to-qanon}.

\begin{figure}
\centering
\includegraphics[width=\columnwidth]{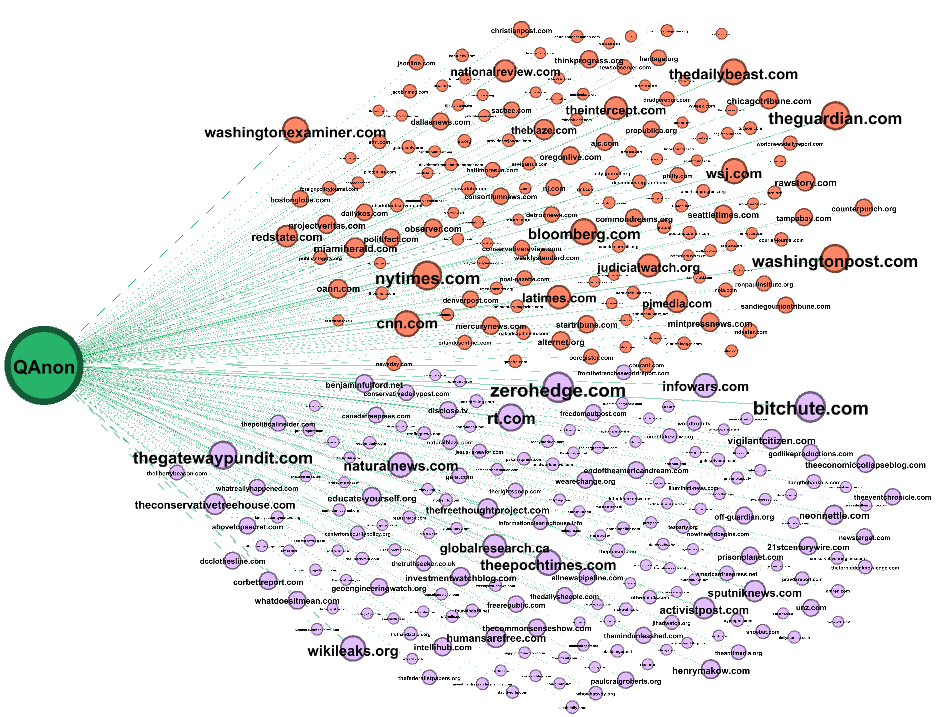}
\caption{\textbf{Websites That QAnon Domains Point To}---  QAnon domains link to both authentic news (orange) and misinformation websites (purple). The size of domain is determined by the number of unique QAnon domains that link to it. QAnon largely points to both misinformation and authentic news website in equal measure. 
}
\label{figure:qanon-to-websites}
\end{figure}
\begin{figure}
\centering
\includegraphics[width=\columnwidth]{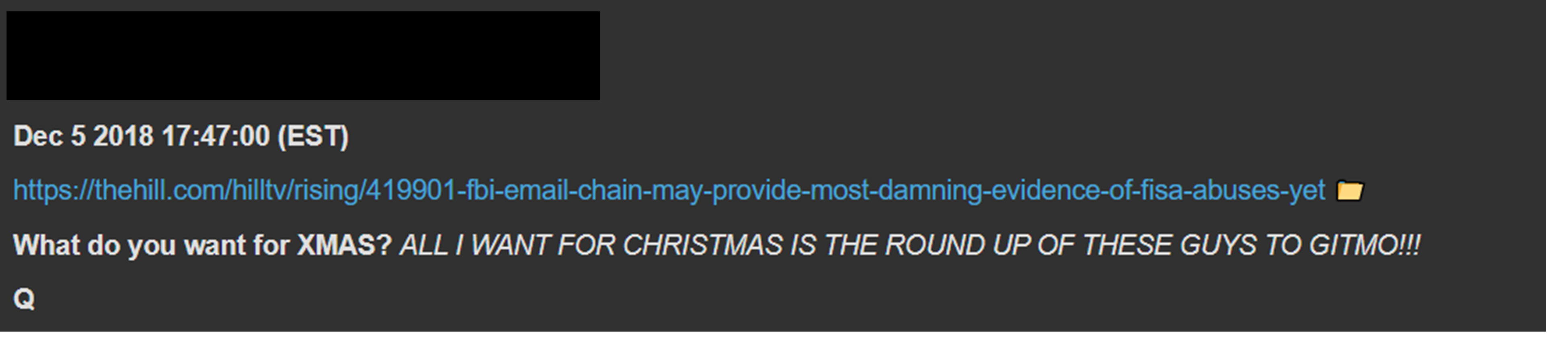}
\caption{\textbf{QAnon Post With Authentic News Article}--- Example from Voat of how QAnon followers make use of authentic news articles. In this case, the supporter is calling for the arrest of political leaders including Hillary Clinton.}
\label{figure:qpost}
\vspace{-10pt}
\end{figure}

In contrast, QAnon sites reference a large number of authentic news sources (Figure~\ref{figure:qanon-to-websites}). For example, in Figure~\ref{figure:qpost}, a QAnon follower references a Hill article to call for the arrest of political leaders like Hillary Clinton. Notably, 148~(88\%) of our authentic news sources are referenced by QAnon sites, with the average QAnon site referencing 26.5~authentic news sources. The most referenced include the New York Times (139~QAnon domains), Washington Post (133~QAnon domains), The Guardian (133~QAnon domains), and Bloomberg (118~QAnon domains). This is consistent with many authentic news outlets being major hubs in the QAnon ecosystem, as was seen in Figure~\ref{fig:tophubs}. 

%
\paragraph{Relationship to Misinformation} As noted previously, {misinformation} and alternative news sources serve as major hubs of QAnon content, regularly linking their readers to QAnon centered websites. We stress that many of these outlets are not new---the Gateway Pundit, Conservative TreeHouse, ZeroHedge, and RT have all been extensively documented for their roles in spreading  misinformation~\cite{starbird2018ecosystem}. 

In contrast to authentic news sources, {misinformation} websites have directed connections to many QAnon supporting domains. After crawling our 189~misinformation domains, we found that 76~(40\%) link to QAnon domains, and these sites link to an average 4.5~QAnon websites. The most prominent websites with directed connections to QAnon are freerepublic.com (90~QAnon domains), benjaminfulford.net (59~QAnon domains), and theconservativetreehouse.com (51~QAnon domains). All four of these websites are known traffickers of conspiracy theories and misinformation~\cite{starbird2018ecosystem}. The difference in the number of directed connections made by misinformation websites with those made by authentic news domains is statistically significant, per a Mann-Whitney U test (Statistic = 12,575, $p<10^{-12}$). We show this difference visually in Figure~\ref{figure:website-to-qanon}.

\begin{figure}
\centering
\includegraphics[width=\columnwidth]{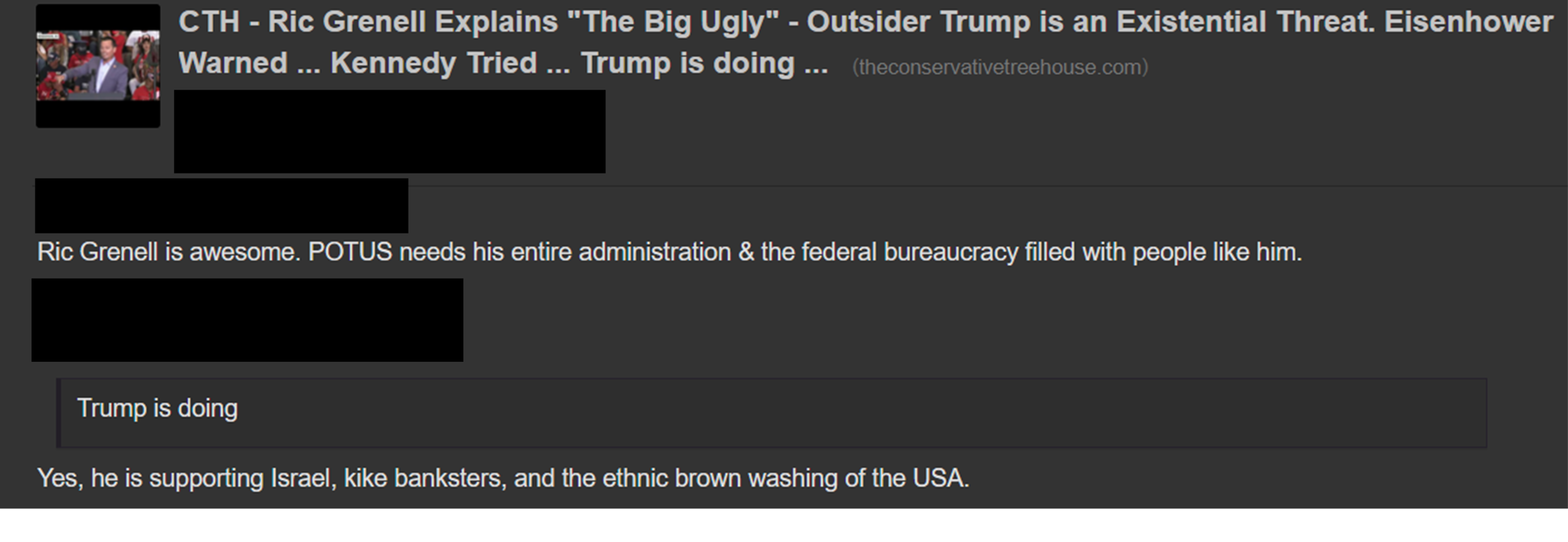}
\caption{\textbf{QAnon Post With Misinformation News Article}--- Example from Voat of how QAnon followers make use of alternative news articles. In this case, the article linked explains why the D.C. ``establishment'' considered former President Trump an ``existential'' threat.}
\label{figure:qpost_Con}
\end{figure}

As with authentic news sources, QAnon sites link to many misinformation domains: 168 of our misinformation domains are referenced by QAnon domains, with an average of 25.1~QAnon domains linking to any given {misinformation} website. QAnon domains frequently link to misinformation content to reinforce their views. For example, the post in Figure~\ref{figure:qpost_Con} shows a QAnon follower criticizing the ``D.C establishment'' while referencing {theconservativetreehouse.com}. The relatively large number of connections from QAnon domains to misinformation sources can be seen in Figure~\ref{figure:qanon-to-websites}.


\paragraph{Correlation in Misinformation Popularity with QAnon Popularity} Given the strong connection between QAnon and {misinformation}, we now consider whether the median popularity of QAnon domains falls and rises with that of {misinformation} websites. We look at the Pearson correlation in popularity of these misinformation sites with our set of QAnon domains over three and a half years (Nov.\ 2017--April 2021). We find that the popularity of QAnon domains and the popularity of our {misinformation} websites have a Pearson correlation of 0.40, substantially higher than between {authentic} news domains and QAnon websites (0.068). This hints that misinformation domains may have played a role in driving users to QAnon content online.

\paragraph{Relationship between QAnon and Popular Websites}
Finally, we measure the relationship between QAnon domains and broadly popular websites, instead of focusing solely only authentic news or misinformation sites. This serves as a verification that the relationship of QAnon websites to {authentic} and {misinformation} news outlets is particularly unique. To do this, we utilize the Alexa top 10K domains from November 1, 2020, checking if QAnon domains link to these websites. We find that QAnon sites link to only 25\% (2506) of these popular sites. On average, only 3~QAnon domains hyperlinked to each website, compared to an average of 25~misinformation and 26~authentic websites. Both misinformation and authentic news domains have a statistically significant difference in hyperlinks with QAnon domains when compared to generally popular websites (Mann-Whitney U test, $p<10^{-12}$). 

These results reinforce that QAnon followers more regularly use news articles (both {authentic} and {misinformation}) to reinforce their arguments compared with other Internet sources. QAnon websites' use of these sources provides a mechanism to further understand both {authentic} and {misinformation} more concretely, which we turn to in the next section.

\section{The QAnon Bellwether: Classifying Misinformation vs.\ Authentic News}

In the last two sections, we showed that QAnon sites have a distinct relationship with news outlets (both authentic and misinformation) compared to other popular Internet domains. In this section, we leverage this insight to use QAnon websites to help identify misinformation websites more broadly.

Recent work by Hounsel et~al.~\cite{hounsel2020identifying} attempted to classify whether websites spread misinformation by training on features derived from a website's domain name, registration, and DNS configuration. However, their fully deployed system is only able to properly identify misinformation with a precision of 5\%~\cite{hounsel2020identifying}. This appears to indicate that more robust features are needed to properly label domains as {misinformation} vs.\ {authentic}. We show that the graph between QAnon sites and other Internet sites can be used to identify whether a domain is spreading {misinformation} with high accuracy and precision.

\subsection{Experimental Setup}
We built a random forest classifier to determine whether a domain is {misinformation} or {authentic news} based on its graph properties and network attributes. We use a random forest model because of its high interpretability. While training, we utilized a randomized hyperparameter search, choosing the models with the highest average accuracy over 100~iterations with five-fold cross-validation. 

\begin{figure}
\centering
\includegraphics[width=\columnwidth]{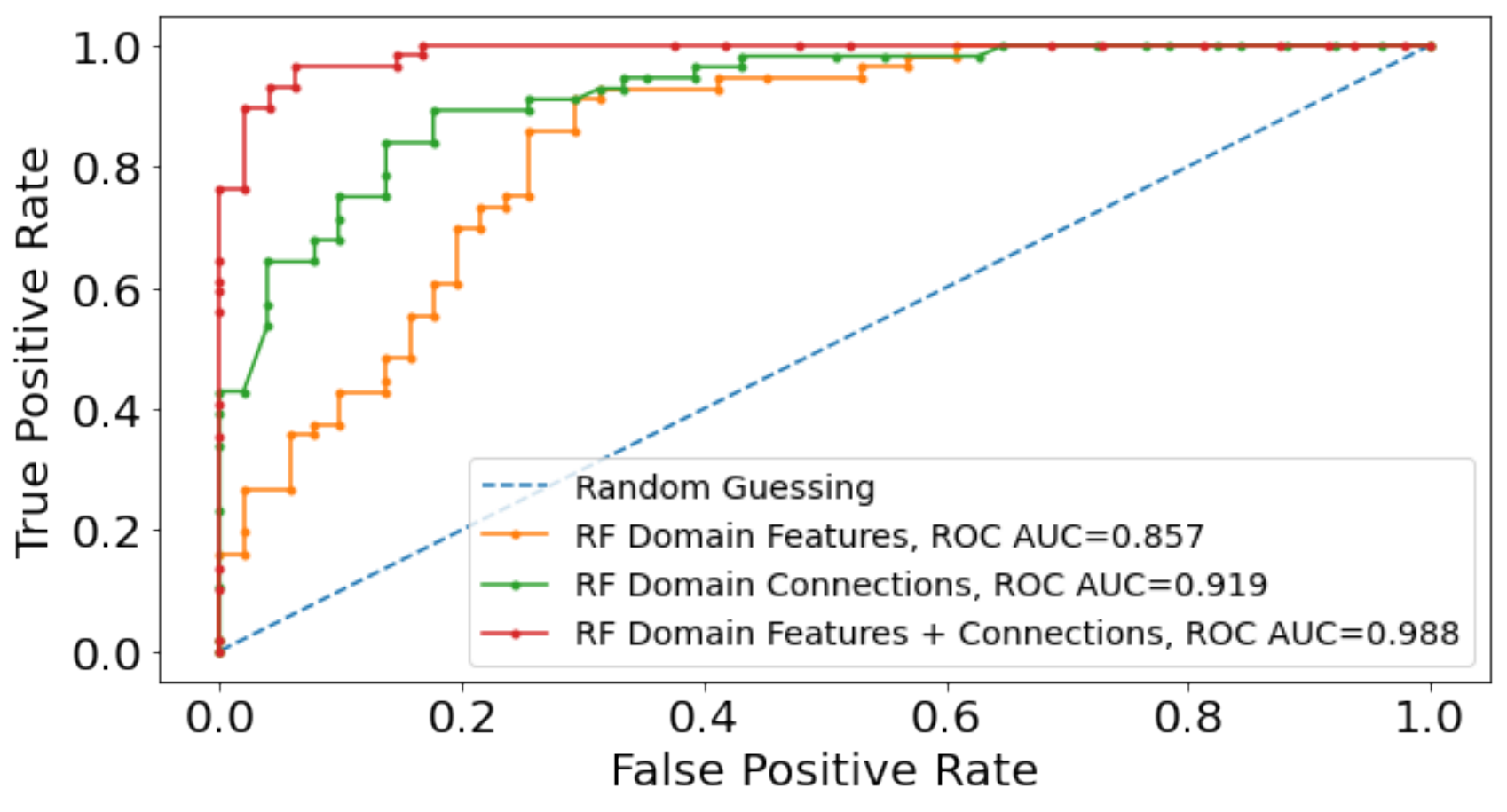} 
\caption{\textbf{Misinformation vs Authentic News Random Forest ROC Curves}--- ROC Curves for the classification of misinformation and authentic news on the 30\% test set. The orange curve (bottom) demonstrates performance of our random forest classifier using only domain features from Hounsel et~al~\cite{hounsel2020identifying}. The green curve (middle) demonstrates performance of our random forest classifier using only the 490 connection features. The red curve (top) demonstrates performance of our random forest classifier using the 490 connection features and domain name features. Note that misinformation domains are a positive instance in our random forest classifier. 
}
\label{figure:classifier}
\end{figure}\looseness=-1

\paragraph{Feature Set} We build a set of binary features derived from whether a given domain has a connection edge (i.e., a hyperlink to) to a set of selected websites. For our set of websites, we utilize the 100~websites that hyperlink to the most QAnon domains, the set of 100~websites that the most QAnon domains link to, and all of the QAnon domains themselves. Removing duplicates, we thus produce 490 binary connection features. As a baseline against which to benchmark our model, we train a random forest model on the domain set features in Hounsel et~al.~\cite{hounsel2020identifying}. These domain features include features based on the domain names, {WHOIS} domain registration data, and autonomous system network numbers. 

\paragraph{Training and Testing Data} We separate our set of 167~authentic and 189~misinformation domains into training and testing data with a 70\%/30\% split. As input to our model, we utilize the domain set features in Hounsel et al.~\cite{hounsel2020identifying} as well as the binary features of whether the set of authentic and misinformation domains have directed edges to the 490 domains in our feature set. 

\subsection{Performance and Evaluation} We train random forest models on three different feature sets in this work: (1) Hounsel et~al.'s domain features, (2) our set of domain connection features, and (3) a combination of Hounsel et~al.'s domain features and our domain connection features. We note for interpretation purposes that we treated misinformation domains as positive instances in our random forest. As seen in Figure~\ref{figure:classifier}, the model using only the QAnon website-derived features performs well, achieving an AUC of 0.919 compared to the model using only domain features, which achieved a mean AUC of 0.843. 
Combining these features leads to even better classification results on the test set. Respectively, these three approaches achieved mean ROC AUCs of 0.8587, 0.919, and 0.988 on the test set. Similarly, we achieved 0.856, 0.928, and 0.980 precision-recall AUCs. These results illustrate the utility of these QAnon domains' gathered features in helping to distinguish authentic/reliable news from misinformation. Because of QAnon's relationship to {misinformation} and conspiratorial websites, other domains' connection to those same websites appears to help identify them as purveyors of conspiracy theories or non-evidenced based misinformation. 

\paragraph{Feature Importance} One key benefit of utilizing the random forest model is the high degree of interpretability. Several of Hounsel et~al.'s features remain important in our combined random forest model. The most important of these features, as measured by the Gini coefficient, is time-info from a website's registration data. Immediately subsequent to these features, we see the autonomous system being important. Like Hounsel et~al.~\cite{hounsel2020identifying}, we note that many of our misinformation domains use consumer-oriented registrars like Namecheap and GoDaddy. 

However, as seen in Figure~\ref{figure:classifier}, our graph connection features play a large role in improving the model. The most important of these connection features are links to social media websites. We also observe that links to prominent misinformation domains are particularly useful, such as links to {globalresearch.ca} and {zerohedge.com}.
As described by Starbird et~al.~\cite{starbird2018ecosystem}, GlobalResearch.ca is a website operated by the Russian-backed Centre for Research on Globalization, and regularly promotes conspiracies and falsehoods. ZeroHedge is a ``far-right libertarian financial blog'' launched in 2009. The site regularly peddles conspiracy theories, most recently those related to COVID-19 vaccines.

We note that our connection features are likely to be robust to changes over time.{Authentic} news outlets rarely link to {misinformation} outlets; this is unlikely to change in the future. Similarly, {misinformation} news websites are likely to still reference extreme or misleading content like QAnon, Covid-19 vaccine skepticism, or election misinformation. For example, a website that continually pushes a conspiracy theory about QAnon will continue to reference QAnon materials on other websites, even if all its registration and other domain features may look similar to that of an authentic news website. For example, we have already shown that many misinformation websites continue to link to QAnon websites. We further note that these features can be regularly updated as QAnon evolves, or new conspiracy theories emerge. Thus, while other features that are used to classify misinformation from authentic news may change over time, these features that are based on conspiracy theories' hyperlink graphs are likely to remain robust.

\section{Discussion}
In this work, we utilize web crawling and hyperlink graphs to uncover and document the structure of QAnon websites as well as to understand the hubs of QAnon content. We further show that using the QAnon hyperlink graph structure, we can identify misinformation sites generally. In this section, we discuss the broader implications of these findings.

\paragraph{QAnon on the Internet}
Many platforms including Google, YouTube, Etsy, Pinterest, Twitch, Discord, Spotify, Vimeo, Patreon, and Twitter have taken steps to curb the spread of the QAnon conspiracy. For example, YouTube,  whenever users search for ``QAnon'' on their platform,  presents a Wikipedia article giving context for the theory. However, despite impressive crackdowns, QAnon material continues to be hyperlinked to by mainstream sources like Twitter, YouTube, PolitiFact, and the New York Times. 
In our analysis, platforms like Twitter and Reddit frequently host content cited by QAnon supporters. Furthermore, these platforms still link to many QAnon associated domains, undercutting their efforts to curb the spread of the conspiracy. QAnon websites are well connected, and once users enter the ecosystem, it is easy for them to discover additional websites that allow the conspiracy theory to spread.

We argue that platforms should consider taking a more proactive role in removing links to known QAnon websites; this may additionally require limiting articles from misinformation and alternative news outlets that continually promote the conspiracy. Some platforms have already begun taking these steps (e.g., Twitter limits the use of BitChute links~\cite{Hinchcliffe2020}), however, much more care should be taken to ensure the conspiracy theory does not continue to grow online. Similarly, personal blogs play a role in promoting QAnon: 17.9\% of the QAnon domains that we documented were hosted by WordPress. Given that WordPress has regularly taken down blogs that espouse the QAnon conspiracy theory or promote violence~\cite{WordpressKick} and given the role the QAnon has had in real-world violence, we also encourage WordPress to consider taking a more proactive role in limiting the promotion of QAnon content.

\paragraph{Studying Conspiracy with Hyperlink Graphs}  Previous work like Starbird et~al.~\cite{starbird2018ecosystem} relies on social media data to characterize misinformation campaigns. Unfortunately, such analyses will likely become increasingly difficult as social media platforms attempt to block the spread of misinformation. In our initial analysis, we found that of the 3,071~Twitter users to linked by the QAnon hotbeds Voat and 8kun, 1,498 (48.8\%) had been banned or had deactivated their account following the attack on the U.S. Capitol on January 6, 2021. At the end of 2020, only 374 (12.1\%) of these users had been banned or had deactivated their accounts, showing how relying on just these social media platforms for research studies can be faulty. Social-media-based approaches will thus largely miss much of the content of conspiracy theories like QAnon. Social media-based analyses also miss the fuller picture of domain hyperlink networks, which can help to illuminate the mutually supporting structures between different types of websites (e.g., news and QAnon).

In contrast, hyperlink graphs are useful for identifying lists of semantically related domains, and thus for compiling lists of QAnon-oriented websites. Similar methods may be utilized to more broadly identify semantically similar conspiracy sites in the future. 

\paragraph{The QAnon Bellwether: Detecting Misinformation}
Our use of QAnon websites themselves as features in our random forest model indicates that these sites may serve as a bellwether for misinformation more broadly. Although the QAnon conspiracy may fade in popularity, we hypothesize that misinformation websites will continue to link to sites related to conspiracy theories and that this can broadly be used as a signal to distinguish misinformation from authentic news. One area of future work is to potentially use the links from misinformation websites as a mechanism to track new conspiracy theories as they appear. Because conspiracy theories like QAnon are much easier to identify due to their extreme beliefs, being able to use them as a signal to label misinformation can serve to largely improve the state of the art. If researchers can proactively identify websites with deep ties to these conspiracy theories and phenomena, researchers may be able to help more effectively stem misinformation and new conspiratorial ideations before they become violent or threaten the public.


\section{Related Work}

Similar to our work, Starbird et~al., built a domain graph of a popular misinformation campaign using links shared on Twitter~\cite{starbird2018ecosystem}. Their approach relies on a Twitter social graph to draw connections between domains, whereas ours relies on the hyperlinks between website pages. Furthermore, Starbird et~al., manage to single out separate out different Russian government agencies and outlets that played prominent roles in helping to spread disinformation, showing explicit coordination between these entities.  Our approach instead maps out the QAnon ecosystem, showing how one can understand the full phenomenon without relying on a social media site. Our approach furthermore is orders of magnitudes larger (3410 links vs. 2.38M websites), largely due to the growth of the QAnon conspiracy theory. We finally note that due to social media crackdowns, accounts that explicitly promote the QAnon conspiracy have been continually removed from the platform, reducing the effectiveness of social media data. For example, top QAnon promoters Lin Wood~\cite{llinwood} and X22Report~\cite{x22report} were removed from Twitter immediately following the attack on the U.S. Capitol on January 6th, 2021. 



As our work measures a specific social phenomenon (the QAnon movement), we also rely on research that has measured other phenomena, for example, attacks on the Syrian White Helmets~\cite{wilson2020cross}, the Manosphere~\cite{ribeiro2020evolution}, disturbing videos targeting children~\cite{papadamou2020disturbed}, and rumors in crisis events~\cite{starbird2018engage}. Most similar to our investigation is from Papasavva et~al.\, who investigate the posting behavior,
emergent themes, and toxicity of the QAnon movement through characterizing the \texttt{/v/GreatAwakening} subcommunity on {voat.co}~\cite{papasavva2020qoincidence}. Their paper shares a similar goal to ours but instead seeks to characterize the content of the QAnon conspiracy, whereas our approach focuses on the spread and development of the theory using links between QAnon websites and other services and web content. Outside of specific investigations, prior work has also studied disinformation networks in the context of images, memes, and toxicity~\cite{zannettou2020characterizing,zannettou2018origins,wang2020understanding,papasavva2020qoincidence}. 

Finally, our work draws on a long history of hyperlink network analysis, going as far back as PageRank~\cite{brin1998anatomy} and early studies that measured the web as a graph~\cite{kleinberg1999web,park2003hyperlink}. In particular, studies show that hyperlink networks to cascade several semantic properties, like sentiment~\cite{miller2011sentiment} and partisanship~\cite{adamic2005political}. Our work extends these prior results to modern online disinformation.

\section{Conclusion}
In this work, we showed that web crawling, hyperlink graphs, and corresponding graph metrics can provide a more holistic understanding of the QAnon conspiracy theory. We showed that hyperlink graphs can be used to find semantically similar websites, which we used to build the largest known set of 324~QAnon websites. We then detailed the structure of the QAnon ecosystem and documented its connections to misinformation websites like BitChute, the Conservative Treehouse, and the Gateway Pundit. We designed a classifier that can differentiate misinformation from authentic news using features built from QAnon websites, showing that knowledge of QAnon link connections can illuminate information about misinformation on the Internet at large. We hope that our methodologies and insight into the QAnon ecosystem can help researchers and online platforms to better study dangerous online conspiracies and misinformation. 


\section{Acknowledgements}
This work was supported in part by the National Science Foundation under grant \#2030859 to the Computing Research Association for the CIFellows Project, a gift from Google, Inc., NSF Graduate Fellowship DGE-1656518, and a Stanford Graduate Fellowship.
\bibliography{paper}

\end{document}